\documentclass[10pt,conference]{IEEEtran}
\IEEEoverridecommandlockouts
\usepackage{cite}
\usepackage{amsmath,amssymb,amsfonts}
\usepackage{physics}
\usepackage{caption}
\usepackage{qcircuit}
\usepackage{algorithm}
\usepackage{algorithmic}
\usepackage{graphicx}
\usepackage{authblk}
\usepackage{textcomp}
\usepackage{xcolor}
\usepackage{subcaption}

\def\BibTeX{{\rm B\kern-.05em{\sc i\kern-.025em b}\kern-.08em
    T\kern-.1667em\lower.7ex\hbox{E}\kern-.125emX}}
\begin{document}
\newcommand{\sol}{MUCIC}
\title{Multi-mode Cavity Centric Architectures for Quantum Simulation}
\author[1]{Samuel Stein\thanks{samuel.stein@pnnl.gov}}
\author[1,3]{Fei Hua}
\author[1]{Chenxu Liu}
\author[2]{Charles Guinn}
\author[1]{James Ang}
\author[3]{Eddy Zhang}
\author[3]{Srivatsan Chakram}
\author[4]{Yufei Ding}
\author[1]{Ang Li}
\affil[1]{Pacific Northwest National Laboratory, Richland, Washington, USA}
\affil[2]{Princeton University, Princeton, New Jersey, USA}
\affil[3]{Rutgers University, New Brunswick, New Jersey, USA}
\affil[4]{University of California San Diego, San Diego, California, USA}

\newcommand{\feicmd}[1] {
  {\color{blue}{#1}}}
    \maketitle
\newcommand{\ang}[1]{{\color{orange}Ang: [#1]}}
\newcommand{\charlie}[1]{{\color{green}Charlie: [#1]}}
\newcommand{\chenxu}[1]{{\color{purple}Chenxu: [#1]}}

\begin{abstract}
Current near-term quantum computing technologies grapple with substantial complexity overheads thereby hindering their ability to meaningfully induce algorithms, necessitating significant engineering and scientific innovations. One class of problems of particular interest is Quantum Simulation, whereby quantum systems are simulated using a quantum computer. However, current devices are yet to surpass classical tensor network techniques. For problems of interest, where classical simulation techniques fail, large degrees of system wide entanglement are required. Furthermore, another challenge of implementing quantum simulation problems is that qubits sit idle whilst alternating simulation terms are implemented, exposing the system to decoherence. In the near term, 2D planar superconducting lattices of circuit-QED elements such as the transmon continue to draw substantial attention, but they are still hindered by their nearest neighbor topology and relatively short lifespan, two problems that are exceedingly problematic for quantum simulation. Arguably the current leading platform, the transmon architecture, has gained substantial attention over the years yet is challenged by these aforementioned two key problems. One emerging technology of particular interest is the multi-mode superconducting resonator capable of storing multiple qubits in one physical device. We observe that these cavities have a natural virtual topology that aligns particularly well with quantum simulation problems, and exhibit much longer lifespans in comparison to other planar superconducting hardware. In this paper we present \sol{} (\underline{MU}lti-mode \underline{C}av\underline{I    }ty \underline{C}entric Architectures for Quantum Simulation), we discuss the relatively simple integration of these devices into the current superconducting quantum computing landscape and their implications to quantum simulation, motivated by their alignment to the quantum simulation problem, and potential promise as a compute-in-memory or quantum memory candidate. We report the development of~\sol{}s transpiler designed for cavity-centric systems, leading to reductions of up to 82\% in quantum simulation circuit depths. Additionally, our investigation demonstrates improvements of up to 19.4\% in converged results from Variational Quantum Algorithms. The insights presented in this paper underscore the potential of superconducting multi-mode cavities in being a strong contender for certain application applications.

\end{abstract}


\section{Introduction}

Quantum Computing continues to dominate headlines, from Quantum Supremacy experiments over random sampling problems \cite{arute2019quantum}, to Quantum Advantage over promising quantum simulation problems \cite{kim2023evidence}. 
These results have served as motivation along the Noisy Intermediate Scale Quantum (NISQ) Era, traversing towards a post-NISQ era. If successful in attaining quantum supremacy, many problems that are currently impossible for classical computing will become tractable. Domains that are poised to change with access to scalable, fault tolerant quantum computing include, but are not limited to, cryptography \cite{shor:siam99}, quantum chemistry \cite{peruzzo+:naturecomm14,Kandala+:Nature17}, and quantum simulation \cite{national2019quantum,dodin2021applications,nielsen2006cluster}.

However, there are still a multitude of hurdles to be overcome in attaining quantum supremacy. On the hardware side, quantum computing contends with complex physical constraints. Accurately controlling quantum systems proves to be exceedingly challenging, with complications propagating across the entire stack. To name a few of these challenges, fabrication of superconducting hardware \cite{de2021materials}, pulse level engineering \cite{de2011second} and calibration, and navigating decoherence via bath interaction~\cite{viola1999dynamical} pose substantial obstacles. Researchers continue to contend with these challenges by improving fabrication procedures such as improved Josephson Junction fabrication \cite{osman2021simplified}, better pulse engineering \cite{gokhale2020optimized}, and optimized bath engineering~\cite{clausen2010bath}, however these problems are still prohibitive. 

\begin{figure*}
    \centering
        \includegraphics[width=1\textwidth]{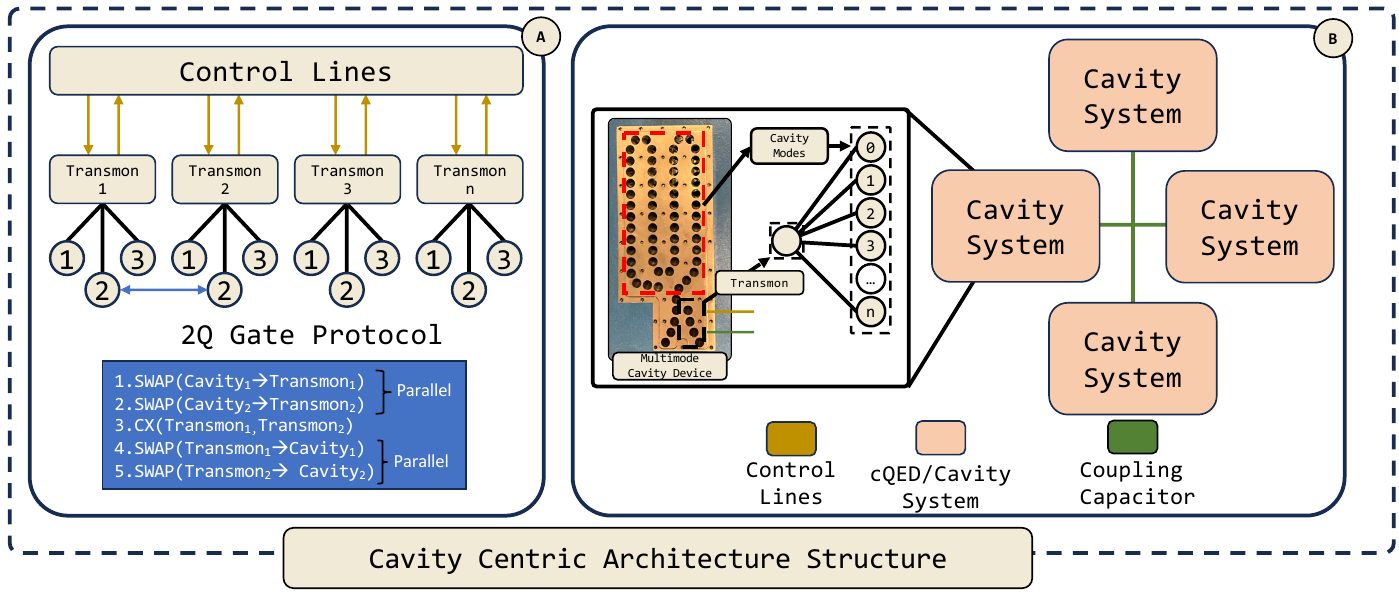}
    \caption{Cavity Centric Architecture Structure example with a Transmon cQED element. (A) represents a possible structure of a Cavity Centric Architecture using a Transmon cQED element. It depicts the control flow required to execute a 2-qubit gate, with control lines representing the necessary connections for controlling and reading out the cQED element. The number of lines per cavity is determined by the requirement for adequate control and readout of the cQED element. (B) demonstrates a potential physical layout for a cavity-centric architecture. Here, transmons are linked either capacitively or through another coupling component. The transmon is situated within the cavity, establishing a capacitive coupling with the various modes of the cavity. Control lines are etched onto the Printed Circuit Board (PCB), which is subsequently inserted into the 3D cavity. This conceptual overview gives an idea of how a cavity-centric architecture may be physically realized and operated.}
    \label{fig:arch_figure}
\end{figure*}

The pursuit of a fault-tolerant quantum computer has spurred significant advancements in superconducting systems, with developments spanning from the transmon \cite{zhang2020high,wallraff2004strong}, fluxonium and Xmon qubits \cite{nguyen2019high,geller2015tunable}. These systems, predominantly championed by IBM and Google, have undergone extensive research, recently achieving a notable transmon $T_1$ time of $503~\mu s$ \cite{wang2021transmon}. Nonetheless, they grapple with inherent physical constraints, such as connectivity issues owing to Purcell decay, frequency collisions, and chip fabrication challenges \cite{smith2022scaling,wang2022towards}, hindering the realization of a fault-tolerant quantum computer. These limitations have given rise to specific architectural topologies such as IBM's Heavy Hex accomodating many of the physical limitations of these architectures. 

The properties of 2D planar superconducting architectures fundamentally misalign with the algorithmic demands of quantum simulation \cite{buluta2009quantum,georgescu2014quantum}. Actual useful quantum simulation demands wide spread entanglement, which is challenging to accomplish in nearest neighbor architectures. To perform just one CNOT gate between two qubits at shortest-distance $N$ away requires at least $3(N-1)$ CNOT gate implementations. Furthermore, sequential application of simulation terms results in consequential idling times, that challenges the relatively short life spans of devices such as the transmon. The inefficiencies in transpiling quantum simulation terms and the propagation of gate-based errors in these architectures underscore the need to explore alternative approaches that navigate challenges such as long range entanglement generation and qubit decoherence during simulation term execution \cite{kim2023evidence}.


Alternative non-planar superconducting quantum devices exist such as superconducting cavities \cite{milul2023superconducting,chakram2021seamless}. In \sol{}, we focus on the \emph{\textbf{multi-mode superconducting cavity}}, and its benefit when designed around as a quantum computing architecture targeting quantum simulation, as it alleviates the key constraints of corresponding 2D planar nearest neighbor topologies and their relatively short lifespan. We focus on specifically the 3D Seamless superconducting cavity \cite{chakram2021seamless}, a 3D cavity manufactured capable of storing multiple microwave qubits in one device through the respective multiple modes, depicted in Figure \ref{fig:photos}. The device's multiple modes are coupled to a transmon, which can be coupled to other qubits. 

Multimode cavities offer a novel architectural paradigm within the superconducting domain, boasting lifespans greatly exceeding 10ms \cite{chakram2021seamless}, a marked improvement over their circuit-QED counterparts like transmon qubits. These cavities present a unique topology that could substantially reduce transpiled coupling distances between qubits compared to the traditional 2D "sea-of-qubits" lattice. However, this system introduces new architectural challenges in transpilation and optimization when targeting quantum simulation such as optimal simulation term partitioning, and scheduling term execution. To prevent substantial routing overhead from naive transpilation, intelligent partitioning strategies, such as k-Way partitioning implemented in \sol{}, are essential. This evident by the cost of an improperly transpiled 2-qubit interaction can cost \textbf{7} time steps, vs the optimal \textbf{3} time steps. To accomplish this improved transpilation, strategies that effectively capitalize on the reduced overhead required for inter-cavity interactions vs the intra-cavity operations counterpart are employed. Despite these complexities, the control mechanisms remain in the microwave regime on the capacitively coupled circuit-QED device, which facilitates a significant reduction in control overhead.

We focus specifically on the problem of Quantum Simulation \sol{}, a realm where today's topologies struggle to tractably handle long-range interactions. \sol{} specifically tackles the current near term problems of quantum simulation by exploiting key features of multi-modal cavities. In doing so, by partitioning and executing simulation terms intelligently, transpilation problems such as $H_4$ can be executed without any rerouting overhead, and with idling simulation terms residing in long-lived cavity modes.

In \sol{}, we demonstrate:
\begin{itemize}
    \item A comprehensive low-level description of the trade offs in circuit-QED element fabrication against cavity fabrication and operation.
    \item A novel transpiler for cavity centric systems targeting quantum simulation, improving circuit depths by up to 82\% on 22 qubit simulation problems when compared with IBM Honeycomb and Rigetti octagonal architectures.
    \item Comprehensive noisy evaluations of cavity centric architectures against their superconducting 2D lattice counterpart demonstrating 19.4\% improvements in Quantum Simulation convergence.
\end{itemize}
\section{Background} \label{sec:background}

\subsection{Quantum Simulation}

In this section, we discuss two key quantum simulation algorithms. Namely, the Variational Quantum Eigensolver and the Quantum Approximate Optimization algorithm. 

\subsubsection{Variational Quantum Eigensolver}
The Variational Quantum Eigensolver (VQE) is a quantum algorithm that integrates classical and quantum computing components, operating based on the variational principle~\cite{baym1990lectures}. The VQE employs a parameterized quantum circuit, referred to as an ansatz, and a Hamiltonian \cite{tang2019} in its operation. This algorithm functions iteratively: initializing the quantum circuit, evaluating the Hamiltonian's energy, and iteratively tuning the parameters to minimize this energy. This process zeroes in on the minimum energy corresponding to the lowest eigenvalue of the Hamiltonian matrix \cite{wiersema2020}. Notably, the VQE has shown promise as a near-term quantum algorithm, particularly for its ability to adapt to temporal noise profiles within the parameterized circuit \cite{fedorov2021}.

Nevertheless, the efficacy of VQEs can be substantially impacted by the choice of ansatz. Particularly in chemical simulations, the Unitary Coupled Cluster Singles and Doubles (UCCSD) ansatz stands as a potent option for achieving chemically precise solutions and is the ansatz we will use in \sol{}. Despite its precision, this ansatz necessitates long-range interactions and poses transpilation challenges \cite{yordanov2020}, with numerous simulation terms forcing non-participating qubits to remain idle, thus increasing the system's susceptibility to decoherence, as well as integrating both local and long range operations, requiring qubit rerouting.

\subsubsection{Quantum Approximate Optimization Algorithm}

The Quantum Approximate Optimization Algorithm (QAOA), as proposed in Farhi et al. \cite{farhi2014quantum}, leverages a parameterized quantum circuit to tackle combinatorial optimization problems. A classic instance of this is the MaxCut problem \cite{majumdar2021optimizing}. In the context of a given graph $G(V,E)$, the MaxCut problem aims to partition $G$ into two subsets $A$ and $B$ in a way that maximizes $C(G) = \sum_{\alpha=1}^{E}G_{E_\alpha}$. Here, $G_{E_\alpha}$ equals 1 if the two vertices are in different subsets (i.e., they are bipartite), and 0 otherwise. 

The QAOA ansatz for a depth-$p$ circuit can be written as $|\psi(\boldsymbol{\gamma}, \boldsymbol{\beta})\rangle = (U(B,\boldsymbol{\beta_i}) U(C,\boldsymbol{\gamma_i}))^p|\psi_{0}\rangle$, where $|\psi_{0}\rangle$ is the initial state, usually chosen as $|+\rangle^{\otimes n}$, and $U(B,\boldsymbol{\beta_i})$ and $U(C,\boldsymbol{\gamma_i})$ are the unitary operators with the respective $i$-th $\beta$ or $\gamma$ value. The unitary operators are defined as $U(C,\boldsymbol{\gamma}) = e^{-i \gamma C} = \prod_{\alpha=1}^{E} e^{-i \gamma G_{E_\alpha}}$ and $U(B,\boldsymbol{\beta}) = e^{-i \beta B} = \prod_{j=1}^{n} e^{-i \beta X_j}$. Where $C$ is the cost Hamiltonian encoding the problem, $B$ is the mixing Hamiltonian, $\boldsymbol{\gamma}$ and $\boldsymbol{\beta}$ are sets of angles, $G_{E_\alpha}$ represents the individual terms in the cost Hamiltonian, and $X_j$ are Pauli-X gates acting on each qubit. The operators $U(C,\boldsymbol{\gamma})$ and $U(B,\boldsymbol{\beta})$ are applied over $p$ layers, where $p$ is the number of times the unitaries are applied. 
    
The goal of the QAOA is to find the optimal parameters $\boldsymbol{\gamma}$ and $\boldsymbol{\beta}$ that minimize the expectation value of the operator $C$, or the cost function $\langle \psi(\boldsymbol{\gamma}, \boldsymbol{\beta}) | C | \psi(\boldsymbol{\gamma}, \boldsymbol{\beta}) \rangle$. 

\subsection{The Challenges of Quantum Simulation}

The implementation of a problem ansatz typically necessitates deep circuits characterized by both short and long-range qubit interactions. For example, in the case of quantum simulation of a hamiltonian approximated via a trotterized Hamiltonians, the problem unitary $e^{-i H \delta t}$ is generated by a Hamiltonian $H = \sum_j H_j$.  Each of these $H_j$ terms break down the problem into implementing individual terms in rapid succession. However, qubits that aren't actively part of a specific $H_j$ term, represented often by an Identity operator in the term. This results in exposing this qubit to decoherence during a simulation time phase that it sits idling for. Furthermore, the group of simulation terms that are to be implemented can require long range interactions, demanding qubit rerouting,  incurring errors. Such problems exert considerable pressure on the hardware's physical topology and strains qubit lifespans adversely.

Current superconducting architecturess do not map well to quantum simulation problems, necessitating the development of architectures that are finely tuned to address these specific challenges, coupled with transpilers optimized to exploit the architectures unique features. \sol{} directly tackles these problems.

\subsection{Superconducting Quantum Systems}

\subsubsection{The Planar Qubit}

Circuit Quantum Electrodynamics (circuit-QED) has emerged as a key framework for studying quantum information processing using superconducting circuits, which form the basis of many present-day quantum computing architectures~\cite{auletta2009quantum}. In such a system, a superconducting LC (Inductor/Capacitance) circuit is etched onto a chip to create a quantum harmonic oscillator. However, alone this is insufficient to act as a qubit. By replacing the inductor with a Josephson Junction, non-linearity is introduced, effectively transforming the quantum harmonic oscillator into an anharmonic oscillator, enabling the behaviour necessary to be considered a qubit. This process creates an 'artificial atom', a fundamental element in circuit-QED based quantum computation \cite{naik2018multimode}. Many realised systems such as transmon qubits \cite{koch2007charge}, fluxonium qubits \cite{nguyen2019high} and $0-\pi$ qubits \cite{gyenis2021experimental}, each offering unique capabilities rely on this circuit QED structure and the Josephson junction. The implementation of the multi modal cavity presented in this paper uses a transmon as the coupling element, however the element choice shares many of the same fundamental challenges of circuit QED based qubits irrespective of specific planar qubit choice. 

One of the most successful implementations in the field of circuit quantum electrodynamics (circuit-QED) is the Transmon qubit, first proposed in \cite{koch2007charge}. This qubit exploits the principle that an increase in the ratio of Josephson energy to charging energy, denoted by $E_J/E_C$, leads to a minor power-law decrease in anharmonicity, while concurrently demonstrating a exponential reduction in charge noise sensitivity. transmon qubits enable rapid gate implementations on the order of $\mathcal{O}(10-100~\text{ns})$, yet these devices are limited by relatively short lifetimes of $\mathcal{O}(0.1-0.5~\text{ms})$ \cite{wei2022hamiltonian,wang2022towards}. The lifetime of a planar transmon is also constrained by dielectric loss, with an upper limit of around $10$~ms \cite{read2023precision}. Scaling transmons presents its own set of challenges. Creating an efficient layout of transmons for a single, large-scale quantum processor has proven to be remarkably difficult. Minor fabrication impurities can significantly impact coherence, and frequency collisions on a large scale for fixed frequency transmons introduce substantial challenges to scaling out \cite{smith2022scaling,murthy2022tof,bilmes2022probing,kim2022effects}. This thus results in two key challenges of the transmon, their \textbf{\emph{limited coupling}} and \textbf{\emph{coherence times}}, and is the underlying reason they are not well suited to quantum simulation problems.

\subsubsection{Hybrid Quantum Systems}

Hybrid Quantum Systems present a compelling approach to advancing quantum computing technology, combining the unique features of diverse quantum systems to exploit their respective strengths \cite{xiang2013hybrid,wang2022ultra}. The key to harnessing the potential of such systems hinges on establishing robust coupling between the constituent quantum systems, such that the computational benefits derived from hybridization outweigh the cost of inter-system interactions. For instance, consider a scenario involving transmon-ion coupling: if a SWAP gate operation requires $1ms$, even though the ion can retain quantum information for durations on the order of seconds, the operation time exceeds the lifespan of the transmon, rendering the hybrid system ineffective.

Developing hybrid systems that combine neutral atoms or trapped ions with superconducting qubits presents considerable challenges due to fundamental differences in control mechanisms and energy scales. Trapped ion systems operate at radio frequencies while neutral atoms and photonic systems typically work in the optical regime. The anharmonicity of these systems often results in weak coupling, complicating interactions with superconducting qubits that operate in the microwave regime. The Planck-Einstein relation, $E=hF$, where F is frequency and h is Planck's constant, underscores this challenge, as it reveals that these systems operate at fundamentally disparate energy levels.

Spatial and control disparities also pose significant obstacles. The physical scales and control techniques employed for superconducting systems and trapped ions or neutral atoms differ drastically. Despite these hurdles, hybrid systems sharing similar energy, control, and scale properties offer the most promising avenues for realizing effective quantum hybridization.

Among these, superconducting hybrid systems demonstrate strong cooperativity, particularly when involving superconducting cavities \cite{braine2023multi} and superconducting qubits. Such systems suggest a promising future for hybrid quantum technologies, as the strengths of superconducting cavities often complement the weaknesses of cQED qubits.

\subsubsection{Superconducting cavities}

Superconducting cavities can host microwave modes, quantizable as harmonic oscillators. However, they inherently lack the non-linearity necessary to function as qubits for executing multi-qubit gates. This necessary non-linearity can be introduced by coupling a circuit-QED element, such as a transmon, to the cavity. This coupling transforms the microwave cavity system into a hybrid quantum system, whereby modes are controlled via dispersive coupling of a circuit-QED element\cite{chakram2021seamless,wang2022high,majumder2022fast}.

Microwave cavities operate in a similar frequency regime to circuit-QED qubits and have high cooperativity, allowing for substantially easier integration \cite{xiang2013hybrid} compared to other hybrid technologies. High Quality Factor (Q-Factor) 3D superconducting cavities show significant promise in terms of extending qubit lifespan, with durations on the order of $\mathcal{O}{(10-100~\text{ms})}$, multiple magnitudes longer than that of transmons \cite{chakram2021seamless}.

However, the fabrication of a 3D cavity presents different challenges compared to the creation of a circuit-QED element. The 3D cavities, being considerably larger, are subject to different loss channels. While transmons present challenges related to microscopic fabrication, the fabrication of 3D cavities encompasses a broader range of difficulties, from microscopic to macroscopic. These challenges include dissipation described by a materials quality factor, dielectric loss from oxide formation, conductor loss, and seam loss \cite{shukla2022quantifying,richardson2016fabrication}, all of which can hinder system performance.

One of the devices greatest strengths is that they are capable of storing multiple qubits in one physical cavity using multiple modes, with a potential of scaling up to $\mathcal{O}(10-1000)$ qubits per cavity being a reasonable estimate \cite{chakram2021seamless}. Another substantial advantage of these systems is the physical control line overhead scales as $\mathcal{O}(1)$ with increasing modes via multiplexing, as these multi-mode cavities are controlled via one single coupled circuit-QED element. 

The ability to store quantum information however is only useful if it can be appropriately interacted with. Universal control of the microwave modes and performing operations in cavity between modes is viable but challenging. Parasitic effect in the form of leakage in both the cavity and transmon, parasitic coupling (comparable to ZZ errors generating varying strengths of undesirable entanglement), and resistance leads to imperfect control and challenges the universal implementation of a high fidelity cavity coupled transmon system performing in-cavity gates. There are two well-known universal instruction sets for microwave cavity mode control: (1) selective number-dependent arbitrary phase (SNAP) gate with displacement operations on the cavity field, where the SNAP gate is defined as
\begin{align}
    \hat{S}(\vec{\phi}) = \sum_{n} e^{i \phi_n} \dyad{n},
\end{align}
where $\ket{n}$ are the corresponding photon number states of the cavity field. The displacement operation is
\begin{align}
    \hat{D}(\alpha) = e^{\alpha a^\dagger - \alpha^* a},
\end{align}
where $\alpha$ is a complex number and $a$ ($a^\dagger$) is the annihilation (creation) operator of the cavity field~\cite{Krastanov2015, Heeres2015}, 
. The (2) second option is the echoed conditional displacement (ECD) gate
\begin{align}
    \text{ECD}(\alpha) = \hat{D}(\frac{\alpha}{2}) \dyad{e}{g} + \hat{D}(-\alpha/2) \dyad{g}{e},
\end{align}
where $\ket{e}$ ($\ket{g}$) is the excited (ground) state of the qubit, and the qubit arbitrary rotations~\cite{Eickbusch2022}. However, due to the slow dispersive coupling between the transmon qubits and the cavity mode, and the relatively low coherence time of the transmons, achieving high-fidelity universal inter-mode control on the microwave modes is still demanding \cite{chakram2021seamless}. Instead of using inter-mode operations, one can SWAP a cavity mode onto the transmon, and then perform the corresponding operations \cite{chakram2021seamless}. 


Despite the unique topological structure and non-homogeneous gate set of superconducting cavities, traditional transpilers such as Paulihedral simply can not operate on these non-homogeneous types of architectures \cite{li2022paulihedral}. One of the primary complexities that the transpiler must facilitate is that only one cavity mode interaction can happen at a time, which essentially restricts a large section of the topology's graph from operation until the current gate operation is completed. Moreover, utilizing the coupled transmon element as the computational portion of the architecture can result in substantial resource overheads in the case of naive transpilation. For example, a process that involves reading the first cavity mode onto the transmon element, transferring it to another idling transmon, then reading out the second mode, followed by the multi-qubit interaction, and finally relocating both qubits back into cavity modes, underlines the costly nature of naive transpilation. This sequence contrasts sharply with the process observed when the modes are hosted disjointedly in their respective cavities. Following Figure \ref{fig:arch_figure}-2Q Gate Protocol's section , the process of simultaneous cavity readouts (Line 1 \& 2), interaction (Line 3), and input back to their original modes (Line 4 \& 5), this can be completed in $3$ time steps. We further want to sequence 2Q Gate Protocols that have an overlap in the respective target qubits, as the sequential SWAP operations commute. This comparison underscores the need to develop a transpiler that can maximize optimal inter-cavity interactions while minimizing routing overhead. When properly implemented, superconducting cavities possess two primary features: a topology characterized by all-to-one coupling ideal for quantum simulation, and long-lived microwave modes reducing the noise-effect of decoherence on idling qubits. We develop \sol{} to exploit these characteristics and transpile quantum simulation problems to cavity centric architectures.

\begin{figure*}
    \centering
    \includegraphics[width=1\textwidth]{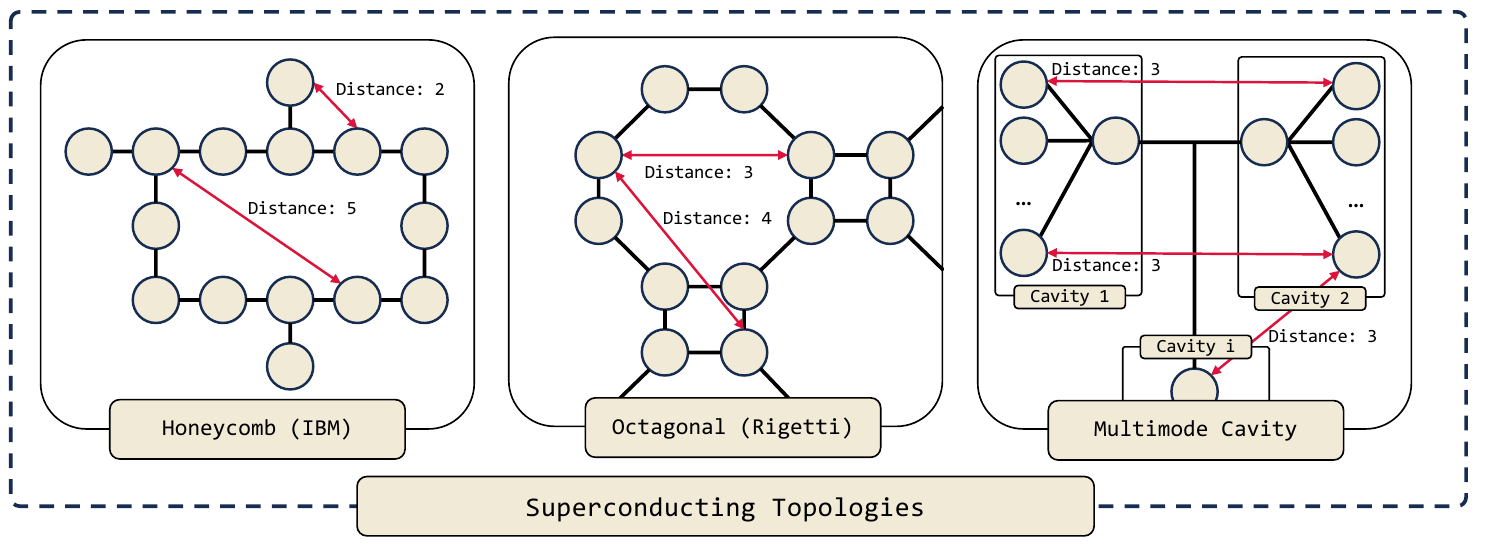}
    \caption{Figure illustrating graph representations of various superconducting platform topologies. The Honeycomb architecture, employed by IBM, and the Octagonal architecture, used by Rigetti, are shown along with a virtual topology representing Cavity-Centric architecture. The Cavity-Centric model involves multiple qubits occupying various cavity modes, each linked to an externally-facing qubit. Inter-qubit communication is facilitated via an external circuit-QED element. The depiction emphasizes the variance in communication cost in 2D-lattice superconducting systems, contrasting with the multi-mode cavity architectures that can maintain constant qubit-qubit distance.}
    \label{fig:topology_figure}
\end{figure*}


\section{Scalable Cavity Centric Architecture}

Current dominating trends in quantum computing architecture comprise IonQ and Quantiniums Trapped Ions \cite{stutz2020trapped,moses2023race,grzesiak2020efficient}, IBM's transmons \cite{gambetta2020ibm}, and QuEra's Rydberg Atoms \cite{adams2019rydberg,wurtz2023aquila}. Current research on cavity architectures discusses their existence and how they could represent a memory \cite{naik2017random}, but lacks how we can think of cavity's in terms of their scalability , topological representation, operation and transpilation. In this section, we address the features that enable scalable cavity centric architectures, and introduce \sol{}. \sol{} integrates the unique features of cavity centric architectures to tackle quantum simulation problems. In this section, we propagate up the quantum computing stack by first discussing the hardwares physical integration and limitations, followed by the abstract topology and its gate set operation.

\subsection{Hardware}

\begin{figure*}
    \centering
    \includegraphics[width=1\textwidth]{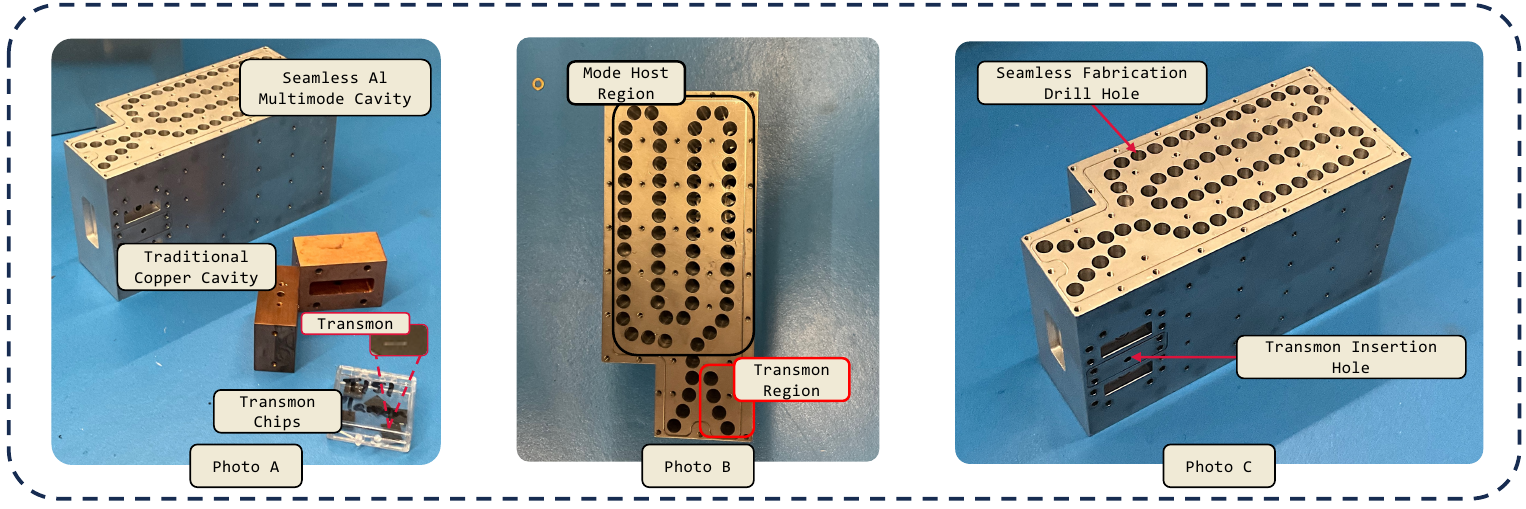}
    \caption{Three photos illustrating an in-house fabricated seamless high-Q Aluminium superconducting multi-mode cavity (photo C, box of silver chips).  Transmons are inserted into the Transmon Insertion Hole and are capacitively coupled to the multiple cavity modes. }
    \label{fig:photos}
\end{figure*}

Cavities, controlled by a non-linear circuit-QED element such as a transmon, are 3D manufactured objects \cite{chakram2021seamless}. The sizes of fabricated cavities are typically within the order of $\mathcal{O}(1-10cm)$, as illustrated in Figure \ref{fig:photos} \cite{chakram2021seamless,milul2023superconducting,burkhart2021error}. This contrasts sharply with transmons, which are on the scale of $\mathcal{O}(100-1000\mu m)$ \cite{wang2022towards,braumuller2016concentric,premkumar2021microscopic}. A close look at the transmon chips box shown in Figure \ref{fig:photos} demonstrates only single-qubit chips, each featuring a small, thin center diagonal line that represents a transmon qubit, highlighted in a red box labeled "Transmon". 

While transmons have a substantially smaller physical footprint - approximately $\mathcal{O}(100\times)$ smaller - they come with their own set of fabrication challenges. These include chip-level issues such as the placement of qubits, which can contribute to Purcell decay and limit the connectivity of transmon \cite{wang2022towards}. Microscopic fabrication complexities also arise, demanding the use of advanced techniques such as photo and electron beam lithography \cite{porrati2022highly,koch2007charge}. Furthermore, the complexity of control line routing necessitates the use of solutions like air-bridges to enable the crossing of superconducting wires \cite{krinner2022realizing,levine2023demonstrating}. Such problems highlight the scaling and fabrication challenges we face as we build more complex circuit-QED architectures.
 
In contrast to superconducting qubits where multiple transmons are fabricated on a single chip, leading to scaling challenges and lower qubit quality on larger chips ~\cite{smith2022scaling}, cavities require only one non-linear cQED element. This element is dipped into the cavity port and coupled to the respective cavity modes via capacitive coupling. Since only one qubit is fabricated on the control chip per cavity, substantially more attention can be devoted to fabricating a single high-quality control element \cite{wang2021transmon}.

The recent invention of the seamless fabrication technique for 3D cavities has mitigated seam loss in cavity operation \cite{patent:20190288367}. These cavities can be fabricated to have a high Q-factor and can undergo iterative post-processing, resulting in higher quality devices \cite{chakram2021seamless}. This advantage stems from the ability to isolate and mill away surface imperfections, which contribute to performance degradation in the cavity. Such an approach mitigates the iterative fabricate-and-test procedure common in superconducting device fabrication.

Regarding the choice of materials, devices constructed from higher Q-factor materials demonstrate improvements in lifespan due to reduced dielectric loss. Current designs typically employ high-purity Aluminium, but there's no reason to believe that future devices made of Niobium wouldn't surpass the lifespans of current cavities.

\subsubsection{Control Challenges}

Planar superconducting qubits typically require on the order of $N$ control lines, such as a flux or charge line, per qubit, resulting in a control complexity that scales linearly with the number of qubits. In contrast, the control complexity for 3D multimode cavities requires only $N$ lines per cavity, with each cavity capable of storing an equivalent of up to $1000$ qubits. This advantage has been demonstrated in \cite{chakram2021seamless}, where a single control element was used to control $10$ modes. The control of the cavity is achieved by manipulating the coupled non-linear cQED element, for example by sideband engineering \cite{chakram2021seamless}. This approach could potentially offer substantial advantages in terms of control complexity compared to other technologies.

\subsubsection{Single Fridge Limitations} 

Superconducting qubits must be operated at temperatures below the superconducting gap of the materials that comprise them and more strictly at low enough temperature such that the the temperature $kT$ is much less than the qubit excitation energies ($<$100 mK). For this reason experiments typically take place on the base plate of a dilution refrigerator.

Superconducting qubits must also be shielded from the noisy microwave environment all around. Typically, devices are kept in superconducting shielded cans $\mathbf{O}(10 cm)$ in diameter bolted to the base plate. Planar devices have additional packaging designed to remove qubit-frequency electromagnetic excitations in the space surrounding the chip.

Typical refrigerator base plates start at 15 cm in diameter \cite{scikro-instru} but larger fridges exist even up to 2 m in diameter \cite{fermilab2022colossal}. Inside of a larger fridge, approximately 20 superconducting shields can fit where each shield either contains one cavity in the cavity-centric architecture or up to a few hundred transmons in planar architectures. 


\subsection{Topology}

One layer above the hardware in the quantum computing stack lies the execution layer, comprised of mapping and transpiling a virtual quantum circuit to a physical device. Cavity-based systems present a unique topological representation, which can be visualized in Figure \ref{fig:topology_figure}. This figure demonstrates that planar qubits are coupled and associated with their respective cavity modes, with each mode storing a single qubits information.

When operating in a hybrid mode, with gates being executed on the circuit-QED qubit, we observe a distinct topological structure. This structure offers a significant topological advantage over the conventional 2D lattice architecture for quantum simulation. In the latter, the distance between any two qubits scales linearly with the number of intermediary transmons. In contrast, in cavity architectures, distance scales sub-linearly because each planar qubit represents a virtually all-to-one connection of $N$ qubits. Regardless of the number of modes, the distance between two modes in two different cavities remains constant, equal to the physical distance between the cavities, with a minimum distance of $3$ as seen in Figure \ref{fig:topology_figure}.

In the case of in-cavity operations, where the coupled cQED element is used not for computation but for performing multi-mode gates, the topological representation changes. Inter-cavity distances of $3$ shift to inter-mode distances of $1$. Each method comes with a challenge: in the hybrid case, communication between qubits within the same cavity becomes challenging, while in the in-cavity operation case, it becomes difficult to communicate between qubits in different cavities. We address this problem in the `Programming \& Transpilation' section by discussing how \sol{} handles this issue through a two-stage transpilation procedure.

\begin{figure*}
    \centering
    \includegraphics[width=1\textwidth]{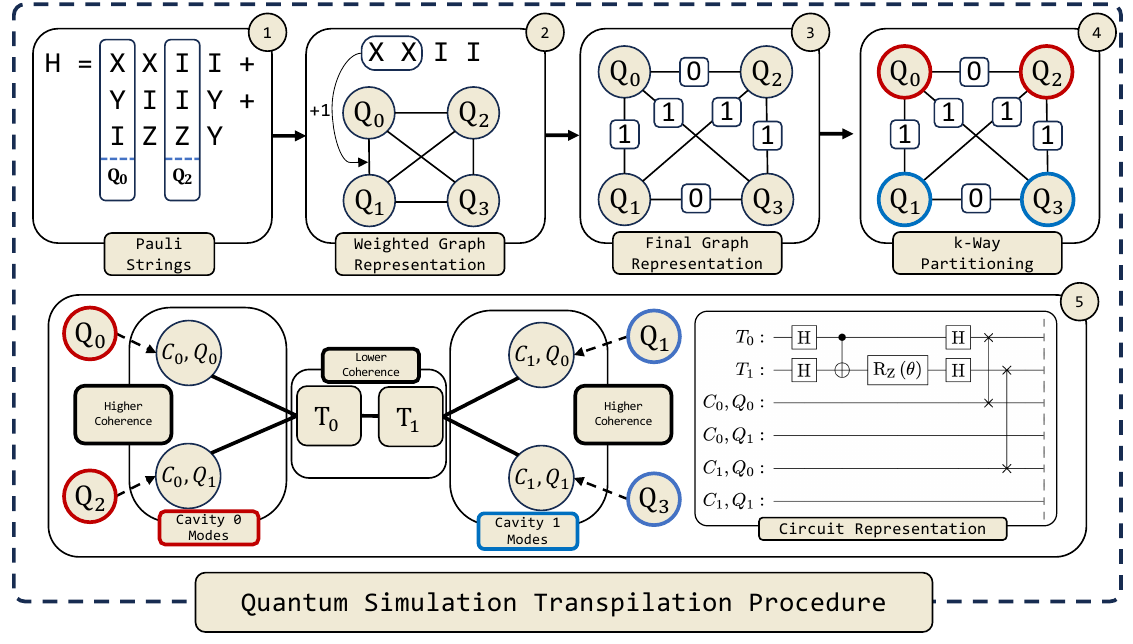}
    \caption{Transpilation Procedure for Quantum Simulation with a hybrid mapping framework. Process depicted in stages, whereby the pauli strings are decomposed into a weighted graph representation of the corresponding Hamiltonian interactions. k-Way partitioning splits the corresponding qubits into k groups. Each partitioned group is placed into to its own separate cavity. A sample XXII operation is represented in the Circuit Representation brick, demonstrating a cavity-transmon circuit representation.}
    \label{fig:cav_centr_arch_transpilation_procedure}
\end{figure*}

\begin{figure}
    \centering
    \includegraphics[width=0.5\textwidth]{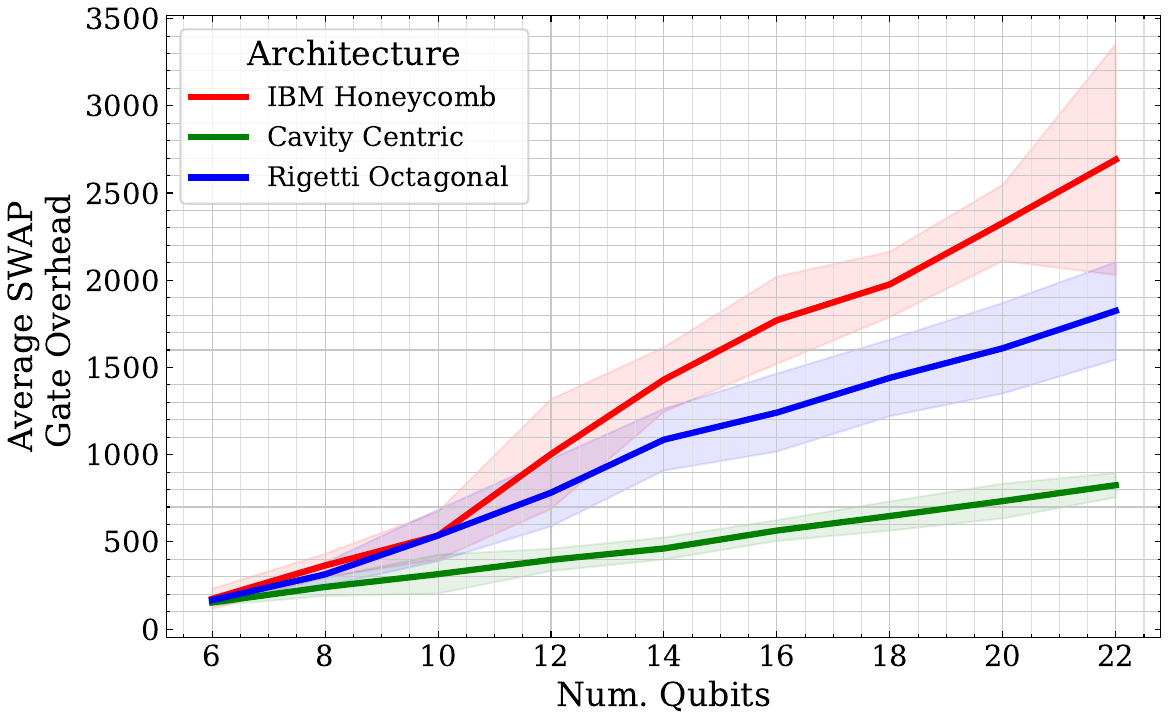}
    \caption{Comparison of routing overhead, calculated via appended routing gates, on IBM and Rigetti Ttransmon systems vs cavity centric systems. 2D centric systems are IBM Honeycomb topologies, as represented in Figure \ref{fig:topology_figure}, and Cavity centric is computed over two cavities, each coupled to a transmon, as represented in Figure \ref{fig:topology_figure}. Number of pauli-terms ranged from $5 \xrightarrow[]{} 30$ for each qubit count, and results were averaged over $~1000$ quantum simulation circuits.}
    \label{fig:routing_overhead}
\end{figure}

\subsection{Operation}

Transmon coupled cavities are capable of driving inter-mode gates via the coupled transmon, however this requires access to all $|f\rangle,|e\rangle,|g\rangle$ levels of the qubits. The operation between two modes requires the photon to spend some time on the transmon, and then be moved to the respective cavity mode, thereby exposing the cavity's long-lived quantum information to the transmons decoherence. As mentioned in Sec.~\ref{sec:background}, a universal control on a cavity mode can be achieved through the coupled transmon. As demonstrated in Ref.~\cite{naik2017random}, entangling gates (CNOT and CZ gates) between two cavity modes can be obtained by swapping one cavity mode state to the transmon and performing quantum gates between the transmon and the other cavity modes. However, this scheme requires pumping the transmon to the second excited state $\ket{f}$ and keeping the excitation on the transmon, which is the most susceptible to loss. In contrast, if cavity modes are used only as memories, only the transmon-cavity SWAP gate is needed. As shown in Refs.~\cite{naik2017random, chakram2021seamless}, iSWAP gates can be implemented using sideband transitions promptly with gate times on the order of $\mathbf{O}(10-100 ns)$.

In the near term, significant efforts are being directed towards optimizing transmons to execute high fidelity gates \cite{noiri2022fast,kjaergaard2020superconducting}. It remains an open question whether transmons or cavities would be more suitable for performing operations, or if inter-mode gates present a superior option. With the current landscape, we envision that transmons will be primarily utilized for gate operations, with cavity modes serving as long-lived storage \cite{naik2017random,stein2023microarchitectures}. Looking further ahead, a compute-in-memory approach may become a viable option. The near term thus represents a hybrid mapping performing inter-cavity operations, transitioning to a cavity-centric mapping performing inter-mode operations in the future. Regardless of the operational mode, a programming and transpilation model is a prerequisite. We focus on hybrid mapping in this paper, however, acknowledge future trends towards in-cavity compute. The transpiler proposed by \sol{} requires very minor adjustments to transition from intra-cavity to inter-cavity operation.

\section{Programming \& Transpilation}

In the quantum computing stack, the transpiler serves as an essential tool, facilitating the decomposition of virtual circuits into hardware-supported basis gates and mapping virtual qubits to physical qubits. This transpiling process presents a greater challenge in cavity-centric hardware due to the distinct gate sets represented by transmon-transmon and cavity-transmon couplings, as illustrated in Figure \ref{fig:topology_figure}. In this section, we introduce the \sol{} transpiler, specifically designed for quantum simulation problems in cavity-centric mappings. As mentioned in \cite{li2022paulihedral}, the flexibility in scheduling within quantum simulation problems allows for order invariance of the each simulation term       in a trotterized time step. This ordering invariance is a crucial feature that enables us to frame a quantum simulation problem as an undirected weighted graph, where the edge weights represent the frequency of qubit interactions. Notably, the absence of time-dependent information on graph edges allows us to decompose the conventional directed graph structure used in most quantum algorithms, where specific prior qubit interactions follow specific ordering. This redefinition enables us to frame our challenge as a partitioning problem, seeking to minimize transpilation overhead by grouping frequently interacting qubits, i.e. those with higher relative weights, into separate physical cavities.

\subsection{Cavity-Transmon Hybrid Mapping of Quantum Simulation}

The challenge of mapping Hamiltonian simulation problems onto 2D lattice architectures arises particularly when dealing with Hamiltonians featuring long-range interactions, given the nearest neighbor coupling in these architectures. This was evident in IBM's recent Advantage experiment \cite{kim2023evidence}, where the targeted Hamiltonian was tailored to interact locally, a key strategy behind quantum tensor networks' spoofing. To demonstrate this, we generate groups of random simulation terms, ranging from 5 terms to 30 terms, over 6 to 22 qubits and transpile to three different architectures. This is demonstrated in Figure \ref{fig:routing_overhead}, where 2D planar lattices demonstrate huge SWAP routing overhead growth as qubit numbers increase. In contrast, the virtual topology of cavity-coupled systems innately allows for efficient execution of both long-range and local interactions, demonstrating up to a $82\%$ reduction in routing overhead, and can be seen by the linear growth pattern of the Cavity Centric architecture. This improvement stems from the sub-linear distance scaling among qubits within cavity architectures coupled with \sol{}'s transpiler, illustrated in Figure \ref{fig:cav_centr_arch_transpilation_procedure}. Furthermore, cavities offer \textbf{\emph{substantially}} longer lifetimes compared to the 2D lattice architectures. This attribute is particularly advantageous in quantum simulation, where a significant number of qubits often remain idle during the physical implementation of terms. The lifespan of circuit QED elements, which is generally shorter, exacerbates the challenges associated with routing. In essence, cavities provide a dual advantage for quantum simulation by addressing two of the most critical issues confronting 2D lattice architectures, namely routing and lifespan constraints.

\begin{figure}
\centering
\begin{subfigure}{.17\textwidth}
\centering
\Qcircuit @C=0.5em @R=0.2em @!R { \\
    \nghost{{q}_{0} :  } & \lstick{{q}_{0} :  } & \ctrl{1} & \qw & \qw \\ 
    \nghost{{q}_{1} :  } & \lstick{{q}_{1} :  } & \targ & \ctrl{2} & \qw \\
    \nghost{{q}_{2} :  } & \lstick{{q}_{2} :  } & \qw & \qw & \qw  \\
    \nghost{{q}_{3} :  } & \lstick{{q}_{3} :  } & \qw & \targ & \ctrl{1} \\ 
    \nghost{{q}_{3} :  } & \lstick{{q}_{4} :  } & \qw & \qw & \targ \\ 
}
\caption{}
\end{subfigure}%
\begin{subfigure}{.17\textwidth}
\centering
\Qcircuit @C=0.5em @R=0.2em @!R { \\
    \nghost{{q}_{0} :  } & \lstick{{q}_{0} :  } & \ctrl{3} & \qw & \qw \\ 
    \nghost{{q}_{1} :  } & \lstick{{q}_{1} :  } & \qw & \ctrl{2} & \qw \\
    \nghost{{q}_{2} :  } & \lstick{{q}_{2} :  } & \qw & \qw & \qw  \\
    \nghost{{q}_{3} :  } & \lstick{{q}_{3} :  } & \targ & \targ & \targ \\ 
    \nghost{{q}_{4} :  } & \lstick{{q}_{4} :  } & \qw & \qw & \ctrl{-1} \\ 
}
\caption{}
\end{subfigure}%
\begin{subfigure}{.17\textwidth}
\centering
\Qcircuit @C=0.5em @R=0.2em @!R { \\
    \nghost{{q}_{0} :  } & \lstick{{q}_{0} :  } & \ctrl{1} & \qw & \qw \\ 
    \nghost{{q}_{1} :  } & \lstick{{q}_{1} :  } & \targ & \ctrl{2} & \qw \\
    \nghost{{q}_{2} :  } & \lstick{{q}_{2} :  } & \qw & \qw & \qw  \\
    \nghost{{q}_{3} :  } & \lstick{{q}_{3} :  } & \qw & \targ & \targ \\ 
    \nghost{{q}_{3} :  } & \lstick{{q}_{4} :  } & \qw & \qw & \ctrl{-1} \\ 
}
\caption{}
\end{subfigure}%
\caption{Circuit (a) represents the entanglement chain structure whereby qubits are entangled in series. (b) represents a root qubit structure whereby qubits are entangled to a target qubit ($q_3$), and (c) represents a mixture of the two.}
\label{fig:updated_circuit}
\end{figure}
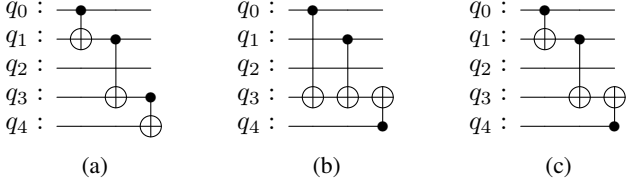

To properly harness the potential of cavity centric architectures, the allocation of qubits into cavity modes requires orchestration. Naive placements results in similar overheads to 2D planar architectures, whereby simulation terms all residing the same cavity demands transmon routing overhead, mitigating the benefit of the cavity centric architecture. To interact any term, the protocol outlined in Figure \ref{fig:cav_centr_arch_transpilation_procedure} is the minimal interaction cost between cavity modes over the transmon. Therefore, we want to maximize this procedure as much as we can, and minimize rerouting terms as best we can. To do this, we consider the structure of a potential virtual quantum simulation circuit, the same as demonstrated in Paulihedral \cite{li2022paulihedral} and visualized in Figure \ref{fig:updated_circuit}.

There are multiple possibilities for entanglement generation: a chain can be established through an arbitrary list of qubits or can entangle qubits to a designated root qubit, or a 
mixed strategy incorporating both mechanisms may be employed. This flexibility affords us an efficient avenue for implementing any quantum simulation term, under the condition that the terms are disjoint by a minimum of one vertex within the cavity-qubit allocation groups. The implementation procedure for any given Pauli string begins with an evaluation of the disjointness among the qubits participating in the simulation term. Should the cumulative vertex difference between any groups be less than two, the entanglement chain can be efficiently implemented. However, if the cumulative difference exceeds two between a cavity group and all other cavity groups, a linear entanglement chain is executed until only pending terms from a single cavity are present. Subsequent to this, the remaining terms are entangled with the root qubit, resulting in a mixed entanglement scheme. Such an approach enables optimal implementation of quantum simulation terms on a cavity centric architecture. Moreover, by grouping and scheduling simulation terms with overlapping interactions, the potential for gate cancellation of SWAPs arises, thus further mitigating the routing overhead.

\begin{algorithm}[!t]
\caption{\sol{} Transpilation Algorithm}
\label{alg:master}
\begin{algorithmic}[1]
\REQUIRE $n_\text{Cavity}$, $n_\text{Modes}$, $n_\text{Sites}$
\STATE $Sim\_Terms \gets \text{[II..X, XI..Y, ..., ZI..X]}$
\STATE $H\_Graph \gets G(V=n_\text{Sites}, E=\text{None})$
\FOR {Term in $Sim\_Terms$}
    \FOR {Qb\_Op in Term}
        \IF{$Operator\ is\ I$}
            \STATE \textbf{continue}
        \ENDIF
        \IF{$Qb\_Prior \neq \text{None}$}
            \STATE $H\_Graph(E=Qb\_Prior,Qb\_Op) += 1$
        \ENDIF
        \STATE{$Qb\_Prior\ is\ Qb\_Op$}
    \ENDFOR
\ENDFOR
\STATE $Groupings \gets n_\text{Cavity}\_Partition(H\_Graph)$
\STATE $E\_Terms$, $L\_Terms \gets Split(Sim\_Terms,Groupings)$
\WHILE{Pending Term}
    \FOR{Term in $E\_Terms$}
        \STATE $Imbalance \gets Balance(Term,Groupings)$
        \IF{$Imbalance < 2$}
            \STATE Transpile(Term) according to Figure \ref{fig:updated_circuit}-A
        \ELSIF{$Imbalance > 2$}
            \STATE Transpile(Term) according to Figure \ref{fig:updated_circuit}-C
        \ENDIF
    \ENDFOR
    \STATE $E\_Terms \gets Reallocate(L\_Terms)$
\ENDWHILE
\end{algorithmic}
\end{algorithm}

In situations where the allocation of qubits to cavities in simulation terms is not disjoint and all belong to a single group, efficient implementation of a simulation term is not feasible. To circumvent this issue, we categorize those terms that cannot be implemented efficiently into distinct groups. Given that the precondition for an efficient execution is disjointness by at least one vertex, we determine the qubit from each cavity involved in the majority of the remaining simulation terms. Each cavity then designates a majority qubit, and a rerouting of qubits among the cavities is executed. This rerouting constitutes the reordering phase of the transpiler. Following this, the simulation proceeds as previously outlined, with the process reiterated until the simulation reaches completion.

We provide a high level description of \sol{} in Algorithm \ref{alg:master}. The algorithm begins by collating a list of commuting simulation terms, stores the number of cavities, their mode counts, and how many sites we're simulating (Line 1-2). A graph is initialized over each site, and weights are populated between nodes with the number of times an interaction is required between the corresponding sites (Line 3-14). A k-Way partitioning algorithm is applied to the graph, where k is the number of cavities we wish to transpile into, and terms that are appropriately allocated are termed $E_{Terms}$, and terms that require rerouting are termed $L_{Terms}$ (Line 15-16).  Pending terms exist if any term in the $E_{Term}$ or $L_{Term}$ group exists.

We demonstrate a concrete example of this transpilation procedure in Figure \ref{fig:cav_centr_arch_transpilation_procedure}. The consistent depth of interactions inherent in cavity-centric architectures is one of the factors that make these architectures particularly well-suited for variational quantum algorithms. Both long-range and short-range interactions come at a fixed cost, given a topology that comprises a modest number of cavities. As the number of cavities increases past where they can be coupled to each other, the inter-cavity routing introduces complexities similar to those of 2D-lattice architectures. However, cavity systems exhibit a sub-linear scaling in routing overhead as overall qubit count increases. The use of an extra set of cavity modes does not introduce any routing overhead. In contrast, 2D-lattice architectures routing overhead scales with the number of qubits on the chip.


\subsection{Alternative Cavity Operation Mapping}

In the case where compute in cavity becomes more viable than the current format, an appropriate transpiler to compensate for this must be engineered. Fortunately, \sol{}s naturally adapts to this transition by pivoting from maximizing group disjointness to minimizing group disjointness. The procedure follows the same methodology, whereby we want to group pauli terms such that cavities contain the maximum number of pauli terms that are implementable without rerouting. The same rerouting computation is done, whereby cavities swap between each other to accommodate terms that could have not been implemented prior. This is iterated upon until no more terms need implementing, and forms the compute-in-memory framework of Quantum Simulation over cavities.


\section{Evaluation}

To evaluate the implications of cavity centric architectures, we simulate quantum simulation problems. Since cavity-centric architectures are relatively new and their exists no publicly accessible cavity based quantum computers,  we choose to simulate results over a noisy coherently limited simulation. We perform density matrix simulation and operate under the assumption that gates are coherence limited. As gates are modeled to be coherently limited, T1/T2 errors are simulated and hence require density matrix simulation. As cavities require simulation of at minimum two extra qubits, the coupled circuit-QED elements, this limits the number of problems we can simulate to relatively modest molecules, however more complex molecules simply demand increased circuit depth and increased interaction lengths, therefore simply placing larger strain on the respective devices. Gate times are set to $100$~ns for a CX gate, or SWAP gates, single qubit gates have time $40$~ns, and I/O gate time of the cavity is $100$~ns. Life spans of transmons are set to $250~\mu$s, and cavity lifespans $30$~ms. Both numbers have been demonstrated in literature \cite{wang2022towards,milul2023superconducting}. We simulate our system using the density matrix simulator of Qiskit 0.44.0 using a server with an Intel Platinum 8276 CPU and 6.4TB main memory. 

\subsection{Variational Quantum Eigensolver}

\begin{figure}
    \centering
    \includegraphics[width=0.5\textwidth]{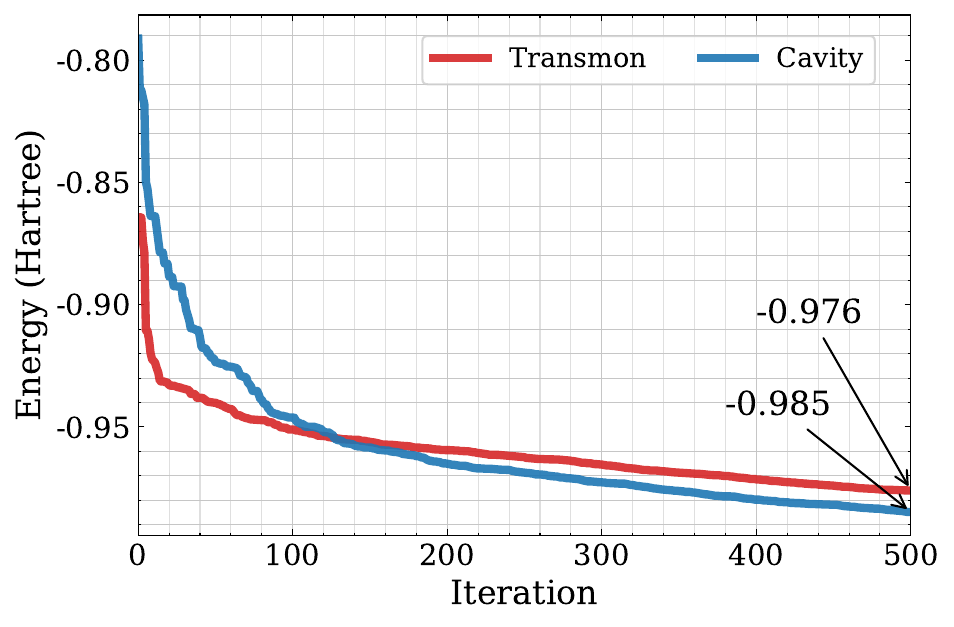}
    \caption{VQE Simulation of H2 comparing Transmon and Cavity relative performance. H2 Simulation comprises 4 qubits. }
    \label{fig:h2_vqe}
\end{figure}

\begin{figure}
    \centering
    \includegraphics[width=0.5\textwidth]{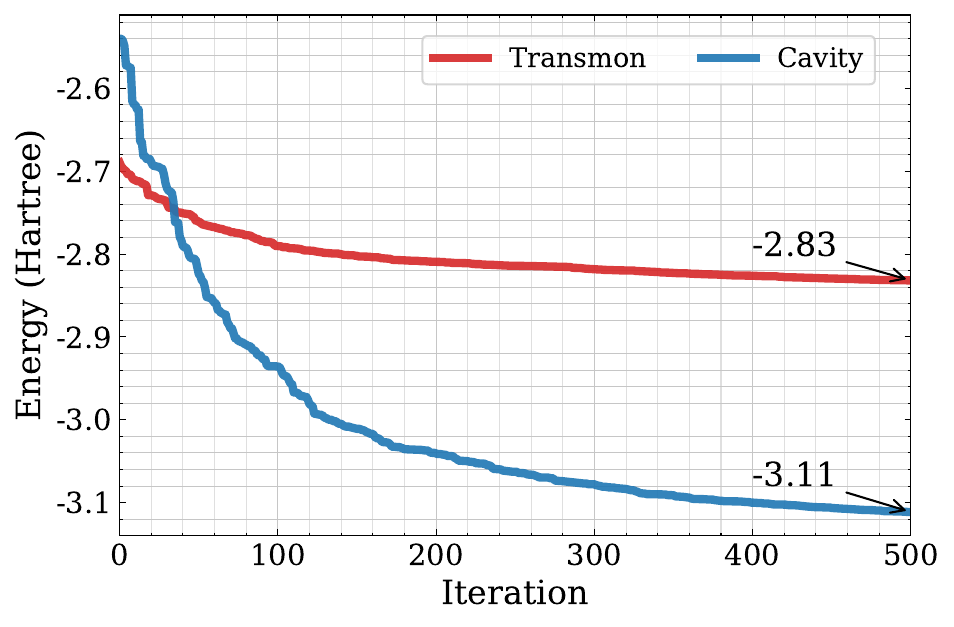}
    \caption{VQE Simulation of H4 comparing Transmon and Cavity relative performance. H4 Simulation comprises 8 qubits.}
    \label{fig:h4_vqe}
\end{figure}

\begin{figure*}
    \centering
    \includegraphics[width=0.9\textwidth]{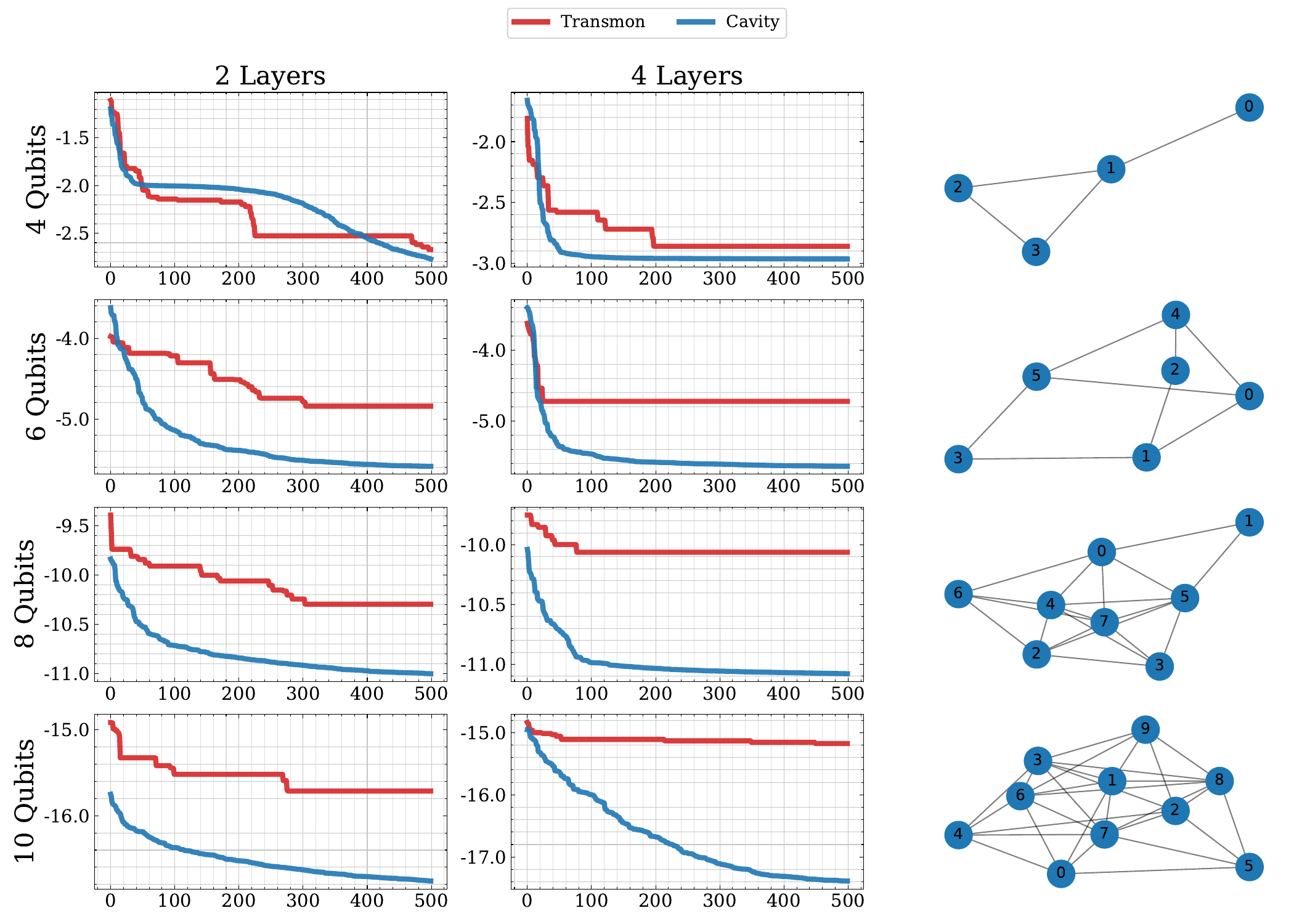}
    \caption{Quantum Approximate Optimization Algorithm results over 4 potential graphs ranging in problem sizes of 4,6,8 and 10 qubits. Comparisons over a 2-layer QAOA ansatz and 4-layer QAOA ansatz are used.}
    \label{fig:QAOA}
\end{figure*}

We simulate molecules H-2 and H-4 using the Unitary Coupled Cluster Single and Double (UCCSD) variational ansatz with a bond length of 1.0\r{A}, in conjunction with the Variational Quantum Eigensolver (VQE), to demonstrate the capabilities of cavity systems in quantum simulations. The qubit requirements for simulating these molecules are 4 for H-2, and 8 for H-4.
In Table \ref{tab:gate_counts}, we focus on the CX and SWAP gate counts of the transpiled versions for each problem, as well as transpile LiH. These transpilations are performed targeting the IBM Manhattan architecture, which features a 2D honeycomb lattice topology. For the cavity systems, we enable SWAP gate transpilation. In real-world applications, the SWAP gate would be tuned according to the Hamiltonian, replacing the role of the CX gate as the primary interaction between the cavity mode and the transmon. The 'SWAP' column captures the fixed I/O overhead inherent to implementing quantum simulations over cavities.
The transmon architecture exhibits a discrepancy in the CX gate count when compared to the cavity system. This difference arises from the routing overhead, which is characteristic of transmon's 2D planar architectures. Although the total gate count appears higher in the cavity systems for these relatively small molecular simulations, cavities offer a distinct advantage: during the routing overhead, idling qubits reside in long-lived cavity modes, thereby being less susceptible to decoherence, as well as having a fixed interaction cost irrespective of the number of qubits in between terms.

\begin{table}[h]
\centering
\caption{Two-Qubit Gate Counts for UCCSD Ansatz transpilation over H-2,H-4 and LiH}
\label{tab:gate_counts}
\begin{tabular}{|c|c|c|c|}
\hline
System & Molecule &  CX  & SWAP  \\
\hline
\hline
Cavity & H2  & 38 & 78 \\
\hline
Transmon & H2 &  62 & - \\
\hline
Cavity & H4  & 1440 & 2880 \\
\hline
Transmon & H4  & 2226 & - \\
\hline\hline
Cavity & LiH  & 8064 & 16128 \\
\hline
Transmon & LiH  & 25457 & - \\
\hline
\end{tabular}
\end{table}

As a result, while the total gate count in cavities on these smaller molecules may be higher, they suffer less from decoherence. Additionally, as we scale to larger simulations, the fixed interaction cost in cavity systems presents another challenge to overcome, especially when compared to lattice systems that lack such fixed costs. Rerouting becomes less of an issue as the number of qubits grows, as only terms that are completely in one cavity require rerouting, resulting in simulations such as H-4 and LiH requiring zero rerouting.
Our transpilation and simulations highlight two primary advantages of using cavity-based architectures for quantum simulations:
\begin{itemize}
\item \textbf{Idling qubits are less susceptible to decoherence during the execution of a time-step in a quantum simulation term.}
\item \textbf{The consistent interaction overhead in cavities enables efficient transpilations. When combined with reduced decoherence, this offers significant benefits for quantum simulations.}
\end{itemize}

We plot the minimum energies obtained from simulation against iterations in Figures \ref{fig:h2_vqe} and \ref{fig:h4_vqe}. As seen in Figure \ref{fig:h2_vqe}, where we simulate the H-2 (Hydrogen) molecule, we see a negligible difference in attained minimum energies with a $>1\%$ difference in converged minimum energies. This can be attributed to the interactions being inherently rather local as well as the circuit depth being relatively short, due to the low qubit count of the simulation. Therefore, the topological constraints of the 2D hardware do not incur overwhelming transpilation overhead. Furthermore, since the longest distance for a qubit to be routed is only 2 positions, since there are only four potential positions and a local interaction only requires two qubits to be next to each other, the routing cost is not overwhelming. 
However, the challenges of 2D planar hardware quickly becomes a problem, as potential qubit allocations grow, long-range entanglement generation is demanded, and idling qubits suffer from rapid decoherence. This can be seen in the H-4 simulation in Figure \ref{fig:h4_vqe}. We observe a rapid convergence for the transmon system with the inability to learn further, whereas the cavity continues to learn and converges to a $9.89\%$ lower energy value. In the transmon case, the 160 quantum simulation terms from the UCCSD ansatz result in a challenging circuit induction over the transmon system. Susceptibility to decoherence and poor topological mappings challenge the transmon case. In the cavity case, long lived cavity modes enable much deeper circuit execution and naturally supported long-range entangling operations mediate arbitrary sized quantum simulation terms. 

\subsection{Quantum Approximate Optimization Algorithm}

To evaluate the performance of \sol{} on the QAOA problem, we generate a set of random graphs and implement the QAOA algorithm over a corresponding cavity centric system, and a transmon based system transpiled to IBM's Honeycomb architecture. Simulation settings are described above under the Evaluation section title. We train over 500 iterations, using the Simultaneous Perturbation Stochastic Approximation (SPSA) optimizer.

The key results for our QAOA simulation are illustrated in Figure \ref{fig:QAOA}. Each row demonstrates a different size graph, and the first column a 2-layer simulation and the second a 4-layer. Values plotted are the lowest cost function evaluation observed over training. The graph in the third column is the problem-graph which we simulate QAOA over. Demonstrated in Figure \ref{fig:QAOA}, smaller circuits of 2-4 layers and 4 qubits demonstrate little to no benefit when using a cavity based system, as similar to the VQE case, interactions are inherently local when operating on fewer qubits. Therefore,  the routing overhead demanded by the transmon system is similar to the cavity based architecture overhead. This is to be expected, with smaller problems not facing substantial routing overhead, and in the case they do it is often over short distances with the largest rerouting distance being 2 in this case. Cavities fixed interaction costs result in a similar performance, as they have much lower routing costs, however require a higher fixed interaction cost. However, as problems get larger, rerouting overheads grow for planar topologies substantially, whilst in the cavity system the routing costs remain fixed. This becomes exaggerated as we grow our problems, evident in the 6-qubit, 8-qubit and 10-qubit plots. This is evident in the growing discrepancies between converged values in Figure \ref{fig:QAOA}. Our relative improvement attained over 2D planar transmon system reaches $3.66$\% for 4-qubits, $19.4$\% for 6-qubits, $10.1$\% for 8-qubits, and $14.6$\% for 10-qubits.

\section{Conclusion}

In this paper we have presented \sol{}, a full stack architecture for implementing quantum simulation over multi-modal superconducting cavities. In this, we highlight the benefits of multi-modal cavity architectures, and demonstrate their natural topological alignment with quantum simulation problems. We utilize a k-way partitioning algorithm over a quantum simulation graph we generate to map to multi modal cavities intelligently, mitigating inter-cavity routing overheads of naive transpilation. Although we focused on quantum simulation in \sol{}, the potential for multi modal cavities propagates forward to the fault tolerant era. Investigations as to how error correcting codes interact with multi modal cavities coupled to two level systems and how threshold is affected are the next steps to understanding the implications of cavities in the NISQ era forward. 
\section*{Acknowledgements}
This material is based upon work supported by the U.S. Department of Energy, Office of Science, National Quantum Information Science Research Centers, Co-design Center for Quantum Advantage (C2QA) under contract number DE-SC0012704, (Basic Energy Sciences, PNNL FWP 76274). The VQE design part was supported by the "Embedding QC into Many-body Frameworks for Strongly Correlated Molecular and Materials Systems'' project, which is funded by the U.S. Department of Energy, Office of Science, Office of Basic Energy Sciences (BES), the Division of Chemical Sciences, Geosciences, and Biosciences (under award 72689). This research used resources of the Oak Ridge Leadership Computing Facility, which is a DOE Office of Science User Facility supported under Contract DE-AC05-00OR22725. This research used resources of the National Energy Research Scientific Computing Center (NERSC), a U.S. Department of Energy Office of Science User Facility located at Lawrence Berkeley National Laboratory, operated under Contract No. DE-AC02-05CH11231.

\cleardoublepage
\bibliographystyle{IEEEtranS}
\bibliography{refs}

\begin{thebibliography}{10}
\providecommand{\url}[1]{#1}
\csname url@samestyle\endcsname
\providecommand{\newblock}{\relax}
\providecommand{\bibinfo}[2]{#2}
\providecommand{\BIBentrySTDinterwordspacing}{\spaceskip=0pt\relax}
\providecommand{\BIBentryALTinterwordstretchfactor}{4}
\providecommand{\BIBentryALTinterwordspacing}{\spaceskip=\fontdimen2\font plus
\BIBentryALTinterwordstretchfactor\fontdimen3\font minus
  \fontdimen4\font\relax}
\providecommand{\BIBforeignlanguage}[2]{{%
\expandafter\ifx\csname l@#1\endcsname\relax
\typeout{** WARNING: IEEEtranS.bst: No hyphenation pattern has been}%
\typeout{** loaded for the language `#1'. Using the pattern for}%
\typeout{** the default language instead.}%
\else
\language=\csname l@#1\endcsname
\fi
#2}}
\providecommand{\BIBdecl}{\relax}
\BIBdecl

\bibitem{scikro-instru}
``Sd system,'' \url{http://www.scikro-instru.com/ProductDetail/3991553.html},
  accessed: 2023-07-10.

\bibitem{adams2019rydberg}
C.~S. Adams, J.~D. Pritchard, and J.~P. Shaffer, ``Rydberg atom quantum
  technologies,'' \emph{Journal of Physics B: Atomic, Molecular and Optical
  Physics}, vol.~53, no.~1, p. 012002, 2019.

\bibitem{arute2019quantum}
F.~Arute, K.~Arya, R.~Babbush, D.~Bacon, J.~C. Bardin, R.~Barends, R.~Biswas,
  S.~Boixo, F.~G. Brandao, D.~A. Buell \emph{et~al.}, ``Quantum supremacy using
  a programmable superconducting processor,'' \emph{Nature}, vol. 574, no.
  7779, pp. 505--510, 2019.

\bibitem{auletta2009quantum}
G.~Auletta, M.~Fortunato, and G.~Parisi, \emph{Quantum mechanics}.\hskip 1em
  plus 0.5em minus 0.4em\relax Cambridge University Press, 2009.

\bibitem{baym1990lectures}
\BIBentryALTinterwordspacing
G.~Baym, \emph{Lectures On Quantum Mechanics}, ser. Advanced book
  program.\hskip 1em plus 0.5em minus 0.4em\relax Avalon Publishing, 1990.
  [Online]. Available: \url{https://books.google.com/books?id=3anvAAAAMAAJ}
\BIBentrySTDinterwordspacing

\bibitem{bilmes2022probing}
A.~Bilmes, S.~Volosheniuk, A.~V. Ustinov, and J.~Lisenfeld, ``Probing defect
  densities at the edges and inside josephson junctions of superconducting
  qubits,'' \emph{npj Quantum Information}, vol.~8, no.~1, p.~24, 2022.

\bibitem{braine2023multi}
T.~Braine, G.~Rybka, A.~Baker, J.~Brodsky, G.~Carosi, N.~Du, N.~Woollett,
  S.~Knirck, M.~Jones, A.~Collaboration \emph{et~al.}, ``Multi-mode analysis of
  surface losses in a superconducting microwave resonator in high magnetic
  fields,'' \emph{Review of Scientific Instruments}, vol.~94, no.~3, 2023.

\bibitem{braumuller2016concentric}
J.~Braum{\"u}ller, M.~Sandberg, M.~R. Vissers, A.~Schneider, S.~Schl{\"o}r,
  L.~Gr{\"u}nhaupt, H.~Rotzinger, M.~Marthaler, A.~Lukashenko, A.~Dieter
  \emph{et~al.}, ``Concentric transmon qubit featuring fast tunability and an
  anisotropic magnetic dipole moment,'' \emph{Applied Physics Letters}, vol.
  108, no.~3, 2016.

\bibitem{buluta2009quantum}
I.~Buluta and F.~Nori, ``Quantum simulators,'' \emph{Science}, vol. 326, no.
  5949, pp. 108--111, 2009.

\bibitem{burkhart2021error}
L.~D. Burkhart, J.~D. Teoh, Y.~Zhang, C.~J. Axline, L.~Frunzio, M.~H. Devoret,
  L.~Jiang, S.~M. Girvin, and R.~J. Schoelkopf, ``Error-detected state transfer
  and entanglement in a superconducting quantum network,'' \emph{PRX Quantum},
  vol.~2, no.~3, p. 030321, 2021.

\bibitem{patent:20190288367}
\BIBentryALTinterwordspacing
D.~S.~N. Chakram, ``Technologies for long-lived 3d multimode microwave
  cavities,'' Patent 20\,190\,288\,367, September, 2019. [Online]. Available:
  \url{https://www.freepatentsonline.com/y2019/0288367.html}
\BIBentrySTDinterwordspacing

\bibitem{chakram2021seamless}
S.~Chakram, A.~E. Oriani, R.~K. Naik, A.~V. Dixit, K.~He, A.~Agrawal, H.~Kwon,
  and D.~I. Schuster, ``Seamless high-q microwave cavities for multimode
  circuit quantum electrodynamics,'' \emph{Physical review letters}, vol. 127,
  no.~10, p. 107701, 2021.

\bibitem{clausen2010bath}
J.~Clausen, G.~Bensky, and G.~Kurizki, ``Bath-optimized minimal-energy
  protection of quantum operations from decoherence,'' \emph{Physical review
  letters}, vol. 104, no.~4, p. 040401, 2010.

\bibitem{de2011second}
P.~de~Fouquieres, S.~G. Schirmer, S.~J. Glaser, and I.~Kuprov, ``Second order
  gradient ascent pulse engineering,'' \emph{Journal of Magnetic Resonance},
  vol. 212, no.~2, pp. 412--417, 2011.

\bibitem{de2021materials}
N.~P. de~Leon, K.~M. Itoh, D.~Kim, K.~K. Mehta, T.~E. Northup, H.~Paik,
  B.~Palmer, N.~Samarth, S.~Sangtawesin, and D.~W. Steuerman, ``Materials
  challenges and opportunities for quantum computing hardware,''
  \emph{Science}, vol. 372, no. 6539, p. eabb2823, 2021.

\bibitem{dodin2021applications}
I.~Y. Dodin and E.~A. Startsev, ``On applications of quantum computing to
  plasma simulations,'' \emph{Physics of Plasmas}, vol.~28, no.~9, p. 092101,
  2021.

\bibitem{Eickbusch2022}
\BIBentryALTinterwordspacing
A.~Eickbusch, V.~Sivak, A.~Z. Ding, S.~S. Elder, S.~R. Jha, J.~Venkatraman,
  B.~Royer, S.~M. Girvin, R.~J. Schoelkopf, and M.~H. Devoret, ``Fast universal
  control of an oscillator with weak dispersive coupling to a qubit,''
  \emph{Nature Physics}, vol.~18, no.~12, pp. 1464--1469, 2022. [Online].
  Available: \url{https://doi.org/10.1038/s41567-022-01776-9}
\BIBentrySTDinterwordspacing

\bibitem{farhi2014quantum}
E.~Farhi, J.~Goldstone, and S.~Gutmann, ``A quantum approximate optimization
  algorithm,'' \emph{arXiv preprint arXiv:1411.4028}, 2014.

\bibitem{fedorov2021}
D.~Fedorov, B.~Peng, N.~Govind, and Y.~Alexeev, ``Vqe method: a short survey
  and recent developments,'' \emph{Materials Theory}, vol.~5, p.~1, 2021.

\bibitem{fermilab2022colossal}
Fermilab, ``It's colossal: Creating the world's largest dilution
  refrigerator,''
  \url{https://news.fnal.gov/2022/12/its-colossal-creating-the-worlds-largest-dilution-refrigerator/},
  2022, accessed: 2023-07-10.

\bibitem{gambetta2020ibm}
J.~Gambetta, ``Ibm’s roadmap for scaling quantum technology,'' \emph{IBM
  Research Blog (September 2020)}, 2020.

\bibitem{geller2015tunable}
M.~R. Geller, E.~Donate, Y.~Chen, M.~T. Fang, N.~Leung, C.~Neill, P.~Roushan,
  and J.~M. Martinis, ``Tunable coupler for superconducting xmon qubits:
  Perturbative nonlinear model,'' \emph{Physical Review A}, vol.~92, no.~1, p.
  012320, 2015.

\bibitem{georgescu2014quantum}
I.~M. Georgescu, S.~Ashhab, and F.~Nori, ``Quantum simulation,'' \emph{Reviews
  of Modern Physics}, vol.~86, no.~1, p. 153, 2014.

\bibitem{gokhale2020optimized}
P.~Gokhale, A.~Javadi-Abhari, N.~Earnest, Y.~Shi, and F.~T. Chong, ``Optimized
  quantum compilation for near-term algorithms with openpulse,'' in \emph{2020
  53rd Annual IEEE/ACM International Symposium on Microarchitecture
  (MICRO)}.\hskip 1em plus 0.5em minus 0.4em\relax IEEE, 2020, pp. 186--200.

\bibitem{grzesiak2020efficient}
N.~Grzesiak, R.~Bl{\"u}mel, K.~Wright, K.~M. Beck, N.~C. Pisenti, M.~Li,
  V.~Chaplin, J.~M. Amini, S.~Debnath, J.-S. Chen \emph{et~al.}, ``Efficient
  arbitrary simultaneously entangling gates on a trapped-ion quantum
  computer,'' \emph{Nature communications}, vol.~11, no.~1, p. 2963, 2020.

\bibitem{gyenis2021experimental}
A.~Gyenis, P.~S. Mundada, A.~Di~Paolo, T.~M. Hazard, X.~You, D.~I. Schuster,
  J.~Koch, A.~Blais, and A.~A. Houck, ``Experimental realization of a protected
  superconducting circuit derived from the 0--$\pi$ qubit,'' \emph{PRX
  Quantum}, vol.~2, no.~1, p. 010339, 2021.

\bibitem{Heeres2015}
\BIBentryALTinterwordspacing
R.~W. Heeres, B.~Vlastakis, E.~Holland, S.~Krastanov, V.~V. Albert, L.~Frunzio,
  L.~Jiang, and R.~J. Schoelkopf, ``Cavity state manipulation using
  photon-number selective phase gates,'' \emph{Phys. Rev. Lett.}, vol. 115, p.
  137002, Sep 2015. [Online]. Available:
  \url{https://link.aps.org/doi/10.1103/PhysRevLett.115.137002}
\BIBentrySTDinterwordspacing

\bibitem{Kandala+:Nature17}
A.~Kandala, A.~Mezzacapo, K.~Temme, M.~Takita, M.~Brink, J.~M. Chow, and J.~M.
  Gambetta, ``{Hardware-efficient variational quantum eigensolver for small
  molecules and quantum magnets},'' \emph{Nature}, vol. 549, no. 7671, pp.
  242--246, 2017.

\bibitem{kim2022effects}
H.~Kim, C.~J{\"u}nger, A.~Morvan, E.~S. Barnard, W.~P. Livingston,
  M.~Alto{\'e}, Y.~Kim, C.~Song, L.~Chen, J.~M. Kreikebaum \emph{et~al.},
  ``Effects of laser-annealing on fixed-frequency superconducting qubits,''
  \emph{Applied Physics Letters}, vol. 121, no.~14, 2022.

\bibitem{kim2023evidence}
Y.~Kim, A.~Eddins, S.~Anand, K.~X. Wei, E.~Van Den~Berg, S.~Rosenblatt,
  H.~Nayfeh, Y.~Wu, M.~Zaletel, K.~Temme \emph{et~al.}, ``Evidence for the
  utility of quantum computing before fault tolerance,'' \emph{Nature}, vol.
  618, no. 7965, pp. 500--505, 2023.

\bibitem{kjaergaard2020superconducting}
M.~Kjaergaard, M.~E. Schwartz, J.~Braum{\"u}ller, P.~Krantz, J.~I.-J. Wang,
  S.~Gustavsson, and W.~D. Oliver, ``Superconducting qubits: Current state of
  play,'' \emph{Annual Review of Condensed Matter Physics}, vol.~11, pp.
  369--395, 2020.

\bibitem{koch2007charge}
J.~Koch, M.~Y. Terri, J.~Gambetta, A.~A. Houck, D.~I. Schuster, J.~Majer,
  A.~Blais, M.~H. Devoret, S.~M. Girvin, and R.~J. Schoelkopf,
  ``Charge-insensitive qubit design derived from the cooper pair box,''
  \emph{Physical Review A}, vol.~76, no.~4, p. 042319, 2007.

\bibitem{Krastanov2015}
\BIBentryALTinterwordspacing
S.~Krastanov, V.~V. Albert, C.~Shen, C.-L. Zou, R.~W. Heeres, B.~Vlastakis,
  R.~J. Schoelkopf, and L.~Jiang, ``Universal control of an oscillator with
  dispersive coupling to a qubit,'' \emph{Phys. Rev. A}, vol.~92, p. 040303,
  Oct 2015. [Online]. Available:
  \url{https://link.aps.org/doi/10.1103/PhysRevA.92.040303}
\BIBentrySTDinterwordspacing

\bibitem{krinner2022realizing}
S.~Krinner, N.~Lacroix, A.~Remm, A.~Di~Paolo, E.~Genois, C.~Leroux,
  C.~Hellings, S.~Lazar, F.~Swiadek, J.~Herrmann \emph{et~al.}, ``Realizing
  repeated quantum error correction in a distance-three surface code,''
  \emph{Nature}, vol. 605, no. 7911, pp. 669--674, 2022.

\bibitem{levine2023demonstrating}
H.~Levine, A.~Haim, J.~S. Hung, N.~Alidoust, M.~Kalaee, L.~DeLorenzo, E.~A.
  Wollack, P.~A. Arriola, A.~Khalajhedayati, Y.~Vaknin \emph{et~al.},
  ``Demonstrating a long-coherence dual-rail erasure qubit using tunable
  transmons,'' \emph{arXiv preprint arXiv:2307.08737}, 2023.

\bibitem{li2022paulihedral}
G.~Li, A.~Wu, Y.~Shi, A.~Javadi-Abhari, Y.~Ding, and Y.~Xie, ``Paulihedral: a
  generalized block-wise compiler optimization framework for quantum simulation
  kernels,'' in \emph{Proceedings of the 27th ACM International Conference on
  Architectural Support for Programming Languages and Operating Systems}, 2022,
  pp. 554--569.

\bibitem{majumdar2021optimizing}
R.~Majumdar, D.~Madan, D.~Bhoumik, D.~Vinayagamurthy, S.~Raghunathan, and
  S.~Sur-Kolay, ``Optimizing ansatz design in qaoa for max-cut,'' \emph{arXiv
  preprint arXiv:2106.02812}, 2021.

\bibitem{majumder2022fast}
S.~Majumder, T.~Bera, R.~Suresh, and V.~Singh, ``A fast tunable 3d-transmon
  architecture for superconducting qubit-based hybrid devices,'' \emph{Journal
  of Low Temperature Physics}, vol. 207, no. 3-4, pp. 210--219, 2022.

\bibitem{milul2023superconducting}
O.~Milul, B.~Guttel, U.~Goldblatt, S.~Hazanov, L.~M. Joshi, D.~Chausovsky,
  N.~Kahn, E.~{\c{C}}ifty{\"u}rek, F.~Lafont, and S.~Rosenblum, ``A
  superconducting quantum memory with tens of milliseconds coherence time,''
  \emph{arXiv preprint arXiv:2302.06442}, 2023.

\bibitem{moses2023race}
S.~Moses, C.~Baldwin, M.~Allman, R.~Ancona, L.~Ascarrunz, C.~Barnes,
  J.~Bartolotta, B.~Bjork, P.~Blanchard, M.~Bohn \emph{et~al.}, ``A race track
  trapped-ion quantum processor,'' \emph{arXiv preprint arXiv:2305.03828},
  2023.

\bibitem{murthy2022tof}
A.~A. Murthy, J.~Lee, C.~Kopas, M.~J. Reagor, A.~P. McFadden, D.~P. Pappas,
  M.~Checchin, A.~Grassellino, and A.~Romanenko, ``Tof-sims analysis of
  decoherence sources in superconducting qubits,'' \emph{Applied Physics
  Letters}, vol. 120, no.~4, p. 044002, 2022.

\bibitem{naik2018multimode}
R.~K. Naik, ``Multimode circuit quantum electrodynamics,''
  \emph{Knowledge@UChicago}, 2018.

\bibitem{naik2017random}
R.~Naik, N.~Leung, S.~Chakram, P.~Groszkowski, Y.~Lu, N.~Earnest, D.~McKay,
  J.~Koch, and D.~I. Schuster, ``Random access quantum information processors
  using multimode circuit quantum electrodynamics,'' \emph{Nature
  communications}, vol.~8, no.~1, p. 1904, 2017.

\bibitem{nguyen2019high}
L.~B. Nguyen, Y.-H. Lin, A.~Somoroff, R.~Mencia, N.~Grabon, and V.~E.
  Manucharyan, ``High-coherence fluxonium qubit,'' \emph{Physical Review X},
  vol.~9, no.~4, p. 041041, 2019.

\bibitem{nielsen2006cluster}
M.~A. Nielsen, ``Cluster-state quantum computation,'' \emph{Reports on
  Mathematical Physics}, vol.~57, no.~1, pp. 147--161, 2006.

\bibitem{noiri2022fast}
A.~Noiri, K.~Takeda, T.~Nakajima, T.~Kobayashi, A.~Sammak, G.~Scappucci, and
  S.~Tarucha, ``Fast universal quantum gate above the fault-tolerance threshold
  in silicon,'' \emph{Nature}, vol. 601, no. 7893, pp. 338--342, 2022.

\bibitem{national2019quantum}
N.~A. of~Sciences~Engineering, Medicine \emph{et~al.}, ``Quantum computing:
  progress and prospects,'' 2019.

\bibitem{osman2021simplified}
A.~Osman, J.~Simon, A.~Bengtsson, S.~Kosen, P.~Krantz, D.~P~Lozano,
  M.~Scigliuzzo, P.~Delsing, J.~Bylander, and A.~Fadavi~Roudsari, ``Simplified
  josephson-junction fabrication process for reproducibly high-performance
  superconducting qubits,'' \emph{Applied Physics Letters}, vol. 118, no.~6,
  2021.

\bibitem{peruzzo+:naturecomm14}
A.~Peruzzo, J.~McClean, P.~Shadbolt, M.-H. Yung, X.-Q. Zhou, P.~J. Love,
  A.~Aspuru-Guzik, and J.~L. O’brien, ``A variational eigenvalue solver on a
  photonic quantum processor,'' \emph{Nature communications}, vol.~5, p. 4213,
  2014.

\bibitem{porrati2022highly}
F.~Porrati, F.~Jungwirth, S.~Barth, G.~C. Gazzadi, S.~Frabboni, O.~V.
  Dobrovolskiy, and M.~Huth, ``Highly-packed proximity-coupled dc-josephson
  junction arrays by a direct-write approach,'' \emph{Advanced Functional
  Materials}, vol.~32, no.~36, p. 2203889, 2022.

\bibitem{premkumar2021microscopic}
A.~Premkumar, C.~Weiland, S.~Hwang, B.~J{\"a}ck, A.~P. Place, I.~Waluyo,
  A.~Hunt, V.~Bisogni, J.~Pelliciari, A.~Barbour \emph{et~al.}, ``Microscopic
  relaxation channels in materials for superconducting qubits,''
  \emph{Communications Materials}, vol.~2, no.~1, p.~72, 2021.

\bibitem{read2023precision}
A.~P. Read, B.~J. Chapman, C.~U. Lei, J.~C. Curtis, S.~Ganjam, L.~Krayzman,
  L.~Frunzio, and R.~J. Schoelkopf, ``Precision measurement of the microwave
  dielectric loss of sapphire in the quantum regime with parts-per-billion
  sensitivity,'' \emph{Physical Review Applied}, vol.~19, no.~3, p. 034064,
  2023.

\bibitem{richardson2016fabrication}
C.~J. Richardson, N.~P. Siwak, J.~Hackley, Z.~K. Keane, J.~E. Robinson,
  B.~Arey, I.~Arslan, and B.~S. Palmer, ``Fabrication artifacts and parallel
  loss channels in metamorphic epitaxial aluminum superconducting resonators,''
  \emph{Superconductor Science and Technology}, vol.~29, no.~6, p. 064003,
  2016.

\bibitem{shor:siam99}
P.~W. Shor, ``Polynomial-time algorithms for prime factorization and discrete
  logarithms on a quantum computer,'' \emph{SIAM review}, vol.~41, no.~2, pp.
  303--332, 1999.

\bibitem{shukla2022quantifying}
P.~Shukla, ``Quantifying the effects of dissipation and temperature on dynamics
  of a superconducting qubit-cavity system,'' \emph{The European Physical
  Journal Plus}, vol. 137, no.~11, p. 1211, 2022.

\bibitem{smith2022scaling}
K.~N. Smith, G.~S. Ravi, J.~M. Baker, and F.~T. Chong, ``Scaling
  superconducting quantum computers with chiplet architectures,'' in \emph{2022
  55th IEEE/ACM International Symposium on Microarchitecture (MICRO)}.\hskip
  1em plus 0.5em minus 0.4em\relax IEEE, 2022, pp. 1092--1109.

\bibitem{stein2023microarchitectures}
S.~Stein, S.~Sussman, T.~Tomesh, C.~Guinn, E.~Tureci, S.~F. Lin, W.~Tang,
  J.~Ang, S.~Chakram, A.~Li \emph{et~al.}, ``Microarchitectures for
  heterogeneous superconducting quantum computers,'' \emph{arXiv preprint
  arXiv:2305.03243}, 2023.

\bibitem{stutz2020trapped}
R.~Stutz, ``Trapped ion quantum computing at honeywell,'' \emph{Bulletin of the
  American Physical Society}, vol.~65, 2020.

\bibitem{tang2019}
H.~L. Tang, V.~O. Shkolnikov, G.~S. Barron, H.~R. Grimsley, N.~Mayhall,
  E.~Barnes, and S.~Economou, ``Qubit-adapt-vqe: An adaptive algorithm for
  constructing hardware-efficient ansätze on a quantum processor,'' \emph{PRX
  Quantum}, vol.~2, p. 020310, 2019.

\bibitem{viola1999dynamical}
L.~Viola, E.~Knill, and S.~Lloyd, ``Dynamical decoupling of open quantum
  systems,'' \emph{Physical Review Letters}, vol.~82, no.~12, p. 2417, 1999.

\bibitem{wallraff2004strong}
A.~Wallraff, D.~I. Schuster, A.~Blais, L.~Frunzio, R.-S. Huang, J.~Majer,
  S.~Kumar, S.~M. Girvin, and R.~J. Schoelkopf, ``Strong coupling of a single
  photon to a superconducting qubit using circuit quantum electrodynamics,''
  \emph{Nature}, vol. 431, no. 7005, pp. 162--167, 2004.

\bibitem{wang2022high}
C.~Wang, I.~Gonin, A.~Grassellino, S.~Kazakov, A.~Romanenko, V.~P. Yakovlev,
  and S.~Zorzetti, ``High-efficiency microwave-optical quantum transduction
  based on a cavity electro-optic superconducting system with long coherence
  time,'' \emph{npj Quantum Information}, vol.~8, no.~1, p. 149, 2022.

\bibitem{wang2021transmon}
C.~Wang, X.~Li, H.~Xu, Z.~Li, J.~Wang, Z.~Yang, Z.~Mi, X.~Liang, T.~Su, C.~Yang
  \emph{et~al.}, ``Transmon qubit with relaxation time exceeding 0.5
  milliseconds,'' \emph{arXiv preprint arXiv:2105.09890}, 2021.

\bibitem{wang2022towards}
------, ``Towards practical quantum computers: Transmon qubit with a lifetime
  approaching 0.5 milliseconds,'' \emph{npj Quantum Information}, vol.~8,
  no.~1, p.~3, 2022.

\bibitem{wang2022ultra}
Z.~Wang, Z.~Bao, Y.~Li, Y.~Wu, W.~Cai, W.~Wang, X.~Han, J.~Wang, Y.~Song,
  L.~Sun \emph{et~al.}, ``An ultra-high gain single-photon transistor in the
  microwave regime,'' \emph{Nature Communications}, vol.~13, no.~1, p. 6104,
  2022.

\bibitem{wei2022hamiltonian}
K.~Wei, E.~Magesan, I.~Lauer, S.~Srinivasan, D.~Bogorin, S.~Carnevale,
  G.~Keefe, Y.~Kim, D.~Klaus, W.~Landers \emph{et~al.}, ``Hamiltonian
  engineering with multicolor drives for fast entangling gates and quantum
  crosstalk cancellation,'' \emph{Physical Review Letters}, vol. 129, no.~6, p.
  060501, 2022.

\bibitem{wiersema2020}
R.~Wiersema, C.~Zhou, Y.~de~Sereville, J.~Carrasquilla, Y.~B. Kim, and H.~Yuen,
  ``Exploring entanglement and optimization within the hamiltonian variational
  ansatz,'' \emph{PRX Quantum}, vol.~1, p. 020319, 2020.

\bibitem{wurtz2023aquila}
J.~Wurtz, A.~Bylinskii, B.~Braverman, J.~Amato-Grill, S.~H. Cantu, F.~Huber,
  A.~Lukin, F.~Liu, P.~Weinberg, J.~Long \emph{et~al.}, ``Aquila: Quera's
  256-qubit neutral-atom quantum computer,'' \emph{arXiv preprint
  arXiv:2306.11727}, 2023.

\bibitem{xiang2013hybrid}
Z.-L. Xiang, S.~Ashhab, J.~You, and F.~Nori, ``Hybrid quantum circuits:
  Superconducting circuits interacting with other quantum systems,''
  \emph{Reviews of Modern Physics}, vol.~85, no.~2, p. 623, 2013.

\bibitem{yordanov2020}
Y.~Yordanov, D.~Arvidsson-Shukur, and C.~Barnes, ``Efficient quantum circuits
  for quantum computational chemistry,'' \emph{Physical Review A}, vol. 102, p.
  062612, 2020.

\bibitem{zhang2020high}
E.~J. Zhang, S.~Srinivasan, N.~Sundaresan, D.~F. Bogorin, Y.~Martin, J.~B.
  Hertzberg, J.~Timmerwilke, E.~J. Pritchett, J.-B. Yau, C.~Wang \emph{et~al.},
  ``High-fidelity superconducting quantum processors via laser-annealing of
  transmon qubits,'' \emph{arXiv preprint arXiv:2012.08475}, 2020.

\end{thebibliography}

\end{document}